\documentclass[sigconf]{acmart}

\setcopyright{none}             
\acmDOI{}                       
\acmISBN{}                      
\acmPrice{}                     

\AtBeginDocument{%
  }

\acmConference[LOCO 2026: 2nd International Workshop on Low Carbon Computing]{}{September 10-11,
  2026}{Lancaster, UK}

\begin{document}

\title{Energy and \texorpdfstring{$\mathrm{CO_2}$}{CO2} Footprint of Climate Model Intercomparison Projects}

\author{Sergi Palomas}
\correspondingauthor
\email{sergi.palomas@bsc.es}
\orcid{0000-0002-2191-152X}
\affiliation{%
  \institution{Barcelona Supercomputing Center (BSC)}
  \city{Barcelona}
  \country{Spain}
}
\affiliation{%
  \institution{Universitat Polit\`ecnica de Catalunya (UPC)}
  \city{Barcelona}
  \country{Spain}
}

\author{Pablo Aparici}
\email{pablo.aparici@bsc.es}
\affiliation{%
  \institution{Barcelona Supercomputing Center (BSC)}
  \city{Barcelona}
  \country{Spain}
}

\author{Gladys Utrera}
\email{gladys.utrera@upc.es}
\affiliation{%
  \institution{Barcelona Supercomputing Center (BSC)}
  \city{Barcelona}
  \country{Spain}
}
\affiliation{%
  \institution{Universitat Polit\`ecnica de Catalunya (UPC)}
  \city{Barcelona}
  \country{Spain}
}

\author{Mario C. Acosta}
\email{mario.acosta@bsc.es}
\affiliation{%
  \institution{Barcelona Supercomputing Center (BSC)}
  \city{Barcelona}
  \country{Spain}
}
\affiliation{%
  \institution{Universitat Polit\`ecnica de Catalunya (UPC)}
  \city{Barcelona}
  \country{Spain}
}

\begin{abstract}
Earth System Models (ESMs) rely heavily on High-Performance Computing (HPC) resources to simulate global climate. As these models evolve, their computational demands continue to grow, driven by three factors: (1) finer spatial grid resolutions, (2) the integration of complex biogeochemical processes (e.g., atmospheric chemistry, interactive vegetation, land use, and ice sheets), and (3) larger climate ensembles to manage uncertainty. Historically, growth in peak computing performance (FLOP/s) has outpaced improvements in energy efficiency (FLOP/Watt), increasing total HPC power consumption. Despite the central role of Model Intercomparison Projects (MIPs) in climate research, quantifying their computational and environmental costs has received limited systematic attention. This paper examines the evolution of climate model carbon accounting from voluntary post-hoc estimation in the Coupled Model Intercomparison Project phase 6 (CMIP6) to standardized accounting under the newly established CMIP7 Task Team on Energy Consumption. Using high-resolution Destination Earth simulations on MareNostrum 5, we empirically evaluate how different accounting boundaries (operational, active-only, and embodied carbon) impact reported energy, carbon emissions, and financial costs. Finally, we outline key methodological considerations for standardizing energy and carbon accounting for Model Intercomparison Projects (MIPs).
\end{abstract}


\maketitle

\section{Introduction}
\begin{figure}[htbp]
    \centering
    \includegraphics[width=0.45\textwidth]{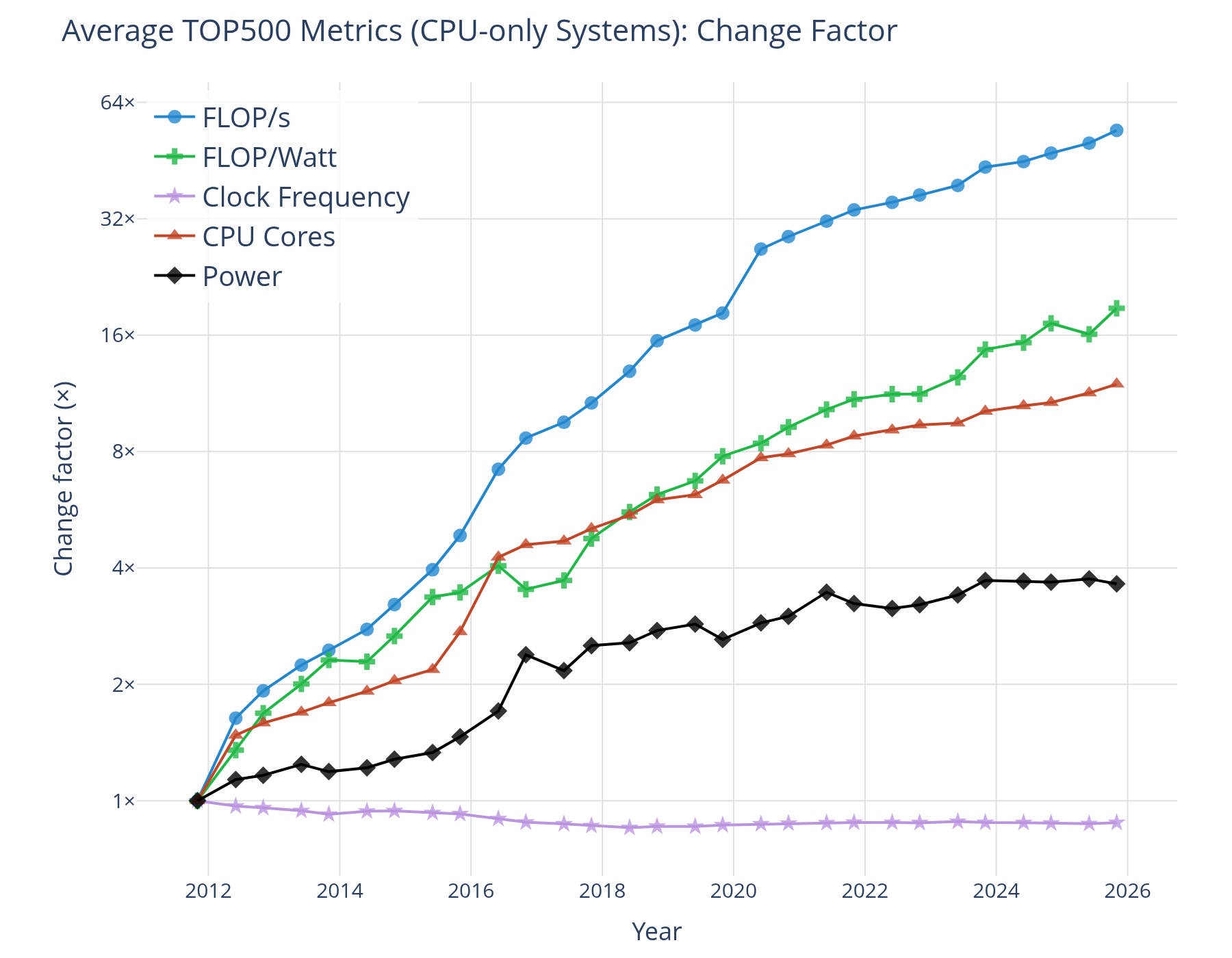}
    \Description{Trends in Top500 supercomputer performance and power draw from 2011 to 2025, demonstrating performance scaling outpacing energy efficiency.}
    \caption{Top500 HPC list (top500.org) averaged change factor from November 2011 to November 2025 (retrieved March 2025).}
    \label{fig:top500_trends}
\end{figure}

High-Performance Computing (HPC) is essential for modern climate modeling. International research consortia, such as the Coupled Model Intercomparison Project (CMIP)~\cite{eyring2016cmip6} coordinated by the World Climate Research Programme (WCRP), rely on large-scale supercomputers to generate climate projections for the Intergovernmental Panel on Climate Change (IPCC). However, power availability and energy efficiency have emerged as critical infrastructure constraints. Between 2011 and 2025, top-tier supercomputers increased peak performance on average by approximately $64\times$, whereas energy efficiency ($\mathrm{FLOP/Watt}$) improved by only $18\times$, driving average facility power draw up to $4\times$ higher (Figure~\ref{fig:top500_trends}). This widening gap between throughput and energy efficiency—often termed the \textit{Power Wall}~\cite{powerwall}—makes energy consumption and carbon emissions critical metrics alongside scientific time-to-solution.

For Earth System Models (ESMs), power and energy limits directly constrain execution at scale. Advancing climate projections requires scaling compute resources across three primary dimensions, each compounding overall energy demand:
\begin{itemize}
    \item \textbf{Spatial Resolution:} Finer grid spacing resolves mesoscale atmospheric and oceanic structures with higher fidelity. However, smaller grid cells require shorter integration time steps, steeply increasing the total floating-point operations per Simulated Year (SY).
    \item \textbf{Model Complexity:} Modern ESMs integrate interactive biogeochemistry, atmospheric chemistry, dynamic vegetation, and ice-sheet dynamics. Coupling these physics sub-models introduces synchronization and communication overheads that can reduce parallel efficiency.
    \item \textbf{Ensemble Scale:} Characterizing scientific uncertainty requires executing multi-model, multi-member ensemble matrices across international modeling centers, multiplying total core-hours and energy consumption.
\end{itemize}

Despite the scale of these computational workloads, energy and carbon accounting in climate science has historically received little attention. However, computational carbon accounting has gained significant traction in machine learning and general High-Performance Computing (HPC). Early carbon estimation in deep learning by Strubell et al.~\cite{strubell-etal-2019-energy} relied on coarse grid averages and static Power Usage Effectiveness (PUE) estimates. Later, Patterson et al.~\cite{patterson2021carbon} showed that using peak chip efficiency overestimates real workload efficiency by $1.6\times$ to $3.5\times$, emphasizing the need for full-rack power measurements. In HPC, Li et al.~\cite{10.1145/3581784.3607035} and Masciari et al.~\cite{10495563} demonstrated that hardware counters (e.g., RAPL, NVML) effectively capture active application power, whereas job schedulers often lack detailed facility PUE integration. Furthermore, studies on dynamic grid carbon intensity~\cite{patterson2021carbon,10.1145/3581784.3607035} revealed that annual average emission factors hide large daily fluctuations, inspiring carbon-aware job scheduling~\cite{10495563}. Beyond operational energy, Wu et al.~\cite{wu2022sustainable} and Li et al.~\cite{10.1145/3581784.3607035} showed that as electricity grids get cleaner, Scope 3 embodied carbon from hardware manufacturing accounts for a large share ($30\%\text{--}50\%+$) of total lifetime emissions. These insights directly inform the carbon accounting choices needed for climate modeling.

To address this gap, the climate modeling community is establishing standardized reporting frameworks. This paper analyzes the transition from voluntary energy reporting in CMIP6 to standardized accounting under the CMIP7 Task Team on Energy Consumption. Using high-resolution Destination Earth simulations on MareNostrum 5, we empirically demonstrate how operational, active-only, and embodied carbon accounting boundaries influence reported energy use, carbon footprint, and financial costs. Finally, we discuss key methodological challenges in standardizing these metrics for Model Intercomparison Projects (MIPs).

\section{Context: Quantifying Model Intercomparison Project Computational Costs}

\subsection{CMIP6 Energy and \texorpdfstring{$\mathrm{CO_2}$}{CO2} Footprint}
During the sixth phase of CMIP (CMIP6)~\cite{eyring2016cmip6}, completed in 2021, 47 participating modeling centers produced petabytes of standardized climate output. A subset of modeling centers voluntarily coordinated under the IS-ENES3 project to collect computational performance metrics and operational costs across diverse HPC platforms~\cite{cmip6-cpmips}. 

To evaluate the environmental impact of CMIP6, Acosta et al.~\cite{cmip6-cpmips} gathered metrics differentiating between \textit{useful} computational execution (simulations producing published scientific data) and \textit{total} execution (encompassing spin-up phases, calibration sweeps, and discarded runs). Carbon Footprints ($\mathrm{t\kern 0.1em CO_2}$) were evaluated by combining useful Simulated Years (SY), facility Power Usage Effectiveness ($\mathrm{PUE}$), and institution-specific energy supplier conversion factors ($\mathrm{kWh} \to \mathrm{CO_2}$).

Across the 8 institutions (out of 45) that reported complete data, the total operational carbon footprint reached $1,692\ \mathrm{t\kern 0.1em CO_2}$.

\begin{figure}[htbp]
    \centering
    \includegraphics[width=0.45\textwidth]{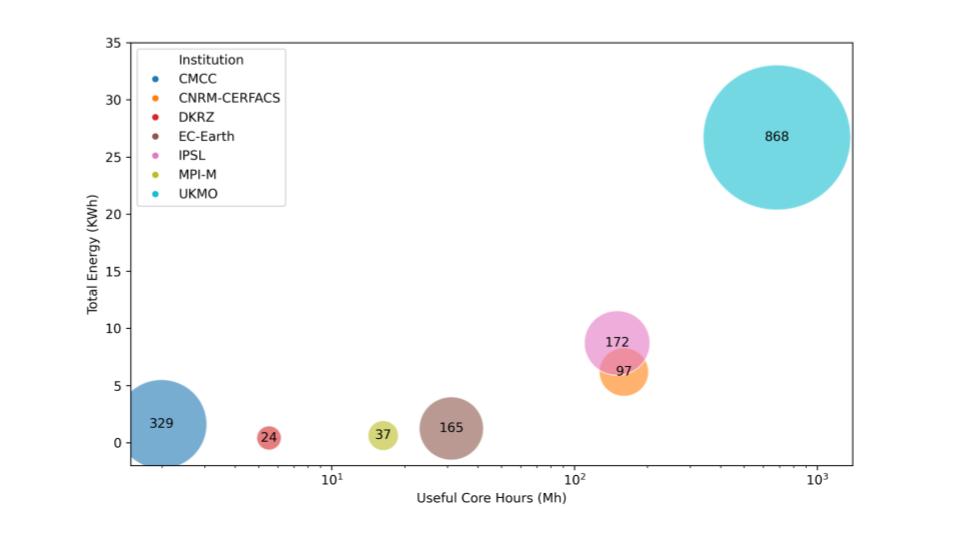}
    \Description{Scatter plot of energy consumption versus core-hours across seven CMIP6 modeling centers with bubble size representing total carbon footprint.}
    \caption{Total Energy vs. Core-Hours across seven tracking modeling centers participating in CMIP6. Bubble size is proportional to total carbon footprint equivalent ($\mathrm{t\kern 0.1em CO_2e}$). Data adapted from Acosta et al. \cite{cmip6-cpmips}.}
    \label{fig:cmip6_footprint}
\end{figure}

Figure~\ref{fig:cmip6_footprint} illustrates total energy consumption, useful core-hours, and carbon footprint ($\mathrm{t\kern 0.1em CO_2e}$) across reporting CMIP6 centers. The data reveal substantial disparities driven by facility efficiency and regional energy grids:
\begin{itemize}
    \item \textbf{Green Tariffs and Grid Carbon Intensity:} Notably, NERC reported a zero operational carbon footprint due to procuring electricity under a 100\% green tariff agreement. Similarly, CERFACS, UKMO, and IPSL benefited from low carbon emissions per kWh from their regional energy suppliers.
    \item \textbf{Facility Power Usage Effectiveness ($\mathrm{PUE}$):} Facility efficiency varied significantly. DKRZ, MPI-M, and NERC reported highly efficient facilities with $\mathrm{PUE} < 1.2$, whereas CERFACS, IPSL, EC-Earth, and UKMO reported significantly higher values. CMCC reported the highest PUE ($1.84$ for the Zeus supercomputer), making it the least power-efficient machine in the collection.
\end{itemize}

\subsection{The CMIP7 Task Team Framework}
\label{sec:cmip7-task-team}
To transition climate modeling from voluntary reporting to systematic carbon accounting, the World Climate Research Programme (WCRP) formally established the \textit{CMIP7 Task Team on Energy Consumption and Carbon Footprint}~\cite{cmip7-taskteam-energy,acosta_2025_17522934}. The task team defines reporting guidelines for all participating MIPs. 

Rather than tracking only published production runs, CMIP7 requests data across production, tuning, and discarded simulations to capture the total computational effort of a modeling campaign~\cite{acosta_2025_17522934}. Carbon footprint accounting adopts the core formulation established during CMIP6~\cite{cmip6-cpmips}, scaling job energy consumption by facility Power Usage Effectiveness ($\mathrm{PUE}$) and site-specific regional grid emission factors:
\begin{equation}
\mathrm{CO_2e} = \mathrm{Energy\ (kWh)} \times \mathrm{PUE} \times \text{Emission Factor}\ (\mathrm{kg\ CO_2e/kWh})
\end{equation}
CMIP7 refines this methodology by prioritizing \textit{dynamic} (active-only) energy consumption—measured via hardware counters (e.g., Intel RAPL, NVIDIA NVML, AMD ROCm-SMI) or scheduler logs—to separate workload power draw from baseline system idle power. The current CMIP7 reporting scope focuses on operational simulation runs on regional power grids, excluding Scope 3 embodied hardware emissions.

\section{Workload Analysis: Climate Digital Twin Simulation Costs}
\label{sec:case-study-cimatedt}
Simulating the Destination Earth model configuration at 5~km resolution on MareNostrum 5 requires 475 compute nodes to achieve a throughput of 1 Simulated Year Per Day ($\mathrm{SYPD}$)~\cite{climatedt}. To evaluate the computational, environmental, and financial costs of this workload, three accounting boundaries can be applied:
\begin{enumerate}
    \item \textbf{Operational Accounting:} Captures full node power draw and facility overhead ($\mathrm{PUE}$) during execution to reflect actual grid electricity demand and Scope~2 emissions.
    \item \textbf{Active-Only Accounting:} Subtracts baseline idle power to isolate dynamic energy consumption, evaluating software and kernel efficiency.
    \item \textbf{Embodied and Life Cycle Accounting:} Extends the boundary to include upstream Scope~3 hardware manufacturing emissions and Total Cost of Ownership ($\mathrm{TCO}$) across the system's lifespan.
\end{enumerate}
The following subsections detail the hardware parameters, data sources, and calculations for each boundary.

\subsection{Operational Accounting}
Operational accounting measures total hardware resource utilization and facility power draw during simulation execution:
\[
475 \text{ nodes} \times 24 \text{ hours} = 11,400 \ \text{node-hours/SY}
\]
Given an empirical average power draw of $900\text{ W}$ per node (measured across over 2,000 baseline runs using the Energy Aware Runtime (EAR) monitoring framework~\cite{ear}), total operational energy consumption is:
\[
11,400 \ \text{node-hours/SY} \times 0.900 \ \mathrm{kWh/node-hour} = 10.3 \ \mathrm{MWh/SY}
\]
Using Spain's 2025 grid carbon intensity ($135\text{ g CO}_2/\text{kWh}$, retrieved from Electricity Maps~\cite{electricitymaps}) and MareNostrum 5's facility Power Usage Effectiveness ($\mathrm{PUE} = 1.08$~\cite{eurohpc-mn5}), the operational Scope~2 carbon footprint is:
\[
1.08 \ (\mathrm{PUE}) \times 10,300 \ \mathrm{kWh/SY} \times 0.135 \ \mathrm{\frac{kg\kern 0.1em CO_2}{kWh}} = 1,501.7 \ \mathrm{kg\kern 0.1em CO_2/SY}
\]
This metric captures the total operational electricity and Scope~2 emissions required to execute the simulation, reflecting real facility utility demand.

\subsection{Active-Only (Dynamic) Accounting}
Active-only accounting isolates dynamic energy and emissions by excluding baseline infrastructure overheads. On MareNostrum 5, an idle compute node draws $390\text{ W}$ (measured via EAR on an allocated node in an idle state over time). Subtracting this baseline yields an active node power draw of $510\text{ W}$ ($900\text{ W} - 390\text{ W}$):
\[
11,400 \ \text{node-hours/SY} \times 0.510 \ \mathrm{kWh/node-hour} = 5.81 \ \mathrm{MWh/SY}
\]
Applying the same PUE and grid intensity, the isolated dynamic Scope~2 carbon footprint is:
\[
1.08 \ (\mathrm{PUE}) \times 5,814 \ \mathrm{kWh/SY} \times 0.135 \ \mathrm{\frac{kg\kern 0.1em CO_2}{kWh}} = 847.7 \ \mathrm{kg\kern 0.1em CO_2/SY}
\]
Crucially, active-only accounting reduces both energy and emissions by $43.5\%$ compared to operational accounting. While this metric isolates software and kernel efficiency for developers, it excludes the idle power needed to keep the cluster available.

\subsection{Embodied Carbon and Total Cost of Ownership Accounting}
Embodied accounting expands the boundary to include upstream Scope~3 emissions and capital costs (CapEx) from hardware manufacturing and procurement. As reported in~\cite{hpc-co2}, the total embodied carbon of MareNostrum 5's General Purpose Partition (GPP) is $22,614\ \mathrm{t\kern 0.1em CO_2e}$. Assuming standard design expectations for the system—a 5-year lifecycle at $80\%$ target utilization—total lifetime capacity across its 7,600 nodes is:
\[
\begin{split}
\text{Total Hours} &= 7,600 \text{ nodes} \times 5 \text{ years} \times 365.25 \text{ days} \\
&\quad \times 24 \text{ hours} \times 0.80 \\
&\approx 266,486,400 \ \text{node-hours}
\end{split}
\]
Dividing embodied carbon by total lifetime node-hours yields the manufacturing carbon intensity:
\[
\begin{split}
\text{Embodied Emissions} &= \frac{22,614,000 \ \mathrm{kg\kern 0.1em CO_2e}}{266,486,400 \ \text{node-hours}} \\
&\approx 0.08486 \ \mathrm{kg\kern 0.1em CO_2e/node-hour}
\end{split}
\]
Multiplying this rate by the $11,400\text{ node-hours}$ required for one simulated year adds an embodied carbon footprint of:
\[
\begin{split}
\text{Embodied Footprint} &= 11,400 \ \text{node-hours/SY} \\
&\quad \times 0.08486 \ \mathrm{kg\kern 0.1em CO_2e/node-hour} \\
&= 967.4 \ \mathrm{kg\kern 0.1em CO_2e/SY}
\end{split}
\]
Combining embodied manufacturing ($967.4\ \mathrm{kg}$) and operational Scope~2 emissions ($1,501.7\ \mathrm{kg}$) yields a total life-cycle carbon footprint of $2,469.1\ \mathrm{kg\kern 0.1em CO_2e/SY}$.
\par
Alongside embodied carbon, financial accounting captures hardware purchase costs plus operational maintenance. Based on the public Total Cost of Ownership (TCO) of €202M for MareNostrum 5's GPP~\cite{eurohpc-mn5} over its 5-year lifecycle at $80\%$ utilization, the TCO per node-hour is:
\[
\begin{split}
\text{TCO/node-hour} &= \frac{202,000,000 \ \text{€}}{
  \left( \begin{array}{c} 
    7,600 \text{ nodes} \times 5 \text{ years} \times 365.25 \text{ days} \\ 
    \times \ 24 \text{ hours} \times 0.80 
  \end{array} \right)
} \\
&\approx 0.76 \ \text{€/node-hour}
\end{split}
\]
This results in a Total Cost of Ownership per simulated year of:
\[
0.76 \ \text{€/node-hour} \times 11,400 \ \text{node-hours/SY} = 8,664 \ \text{€/SY}
\]
This metric includes the upfront capital investment (CapEx) required to build the supercomputer alongside ongoing operational costs, showing that hardware purchase is a primary cost driver.

\section{Challenges and Discussion}

\subsection{Comparative Analysis of Accounting Boundaries}
The empirical results in Section~\ref{sec:case-study-cimatedt} illustrate how the choice of accounting boundary alters reported metrics across energy, carbon, and financial dimensions:
\begin{itemize}
    \item \textbf{Active-Only Boundary (Dynamic Workload):} Isolates dynamic power ($510\text{ W}$), energy ($5.81\text{ MWh/SY}$), and Scope~2 carbon ($847.7\ \mathrm{kg\ CO_2e/SY}$) for algorithm and kernel optimization. While valuable for software developers, it underestimates facility energy consumption and real emissions by $43.5\%$ by omitting baseline idle power ($390\text{ W/node}$).
    \item \textbf{Operational Boundary (Facility Allocation):} Captures full node electricity consumption ($10.3\text{ MWh/SY}$) and total operational Scope~2 carbon ($1,501.7\ \mathrm{kg\ CO_2e/SY}$). It reflects true datacenter utility demand and grid emissions, but remains sensitive to regional grid carbon intensity and omits hardware manufacturing debt.
    \item \textbf{Embodied and Life Cycle Boundary (Full System Debt):} Integrates Scope~3 manufacturing debt ($967.4\ \mathrm{kg\ CO_2e/SY}$), resulting in a $1.64\times$ multiplier over operational carbon (totaling $2,469.1\ \mathrm{kg\ CO_2e/SY}$), and accounts for the Total Cost of Ownership (€8,664/SY, based on €202M in system procurement). This comprehensive boundary prevents carbon leakage and provides full accountability for institutional planning.
\end{itemize}

\subsection{Methodological Challenges}
\subsubsection{Measurement Accuracy vs. Usability}
Accurately profiling energy across heterogeneous HPC clusters while maintaining cross-platform comparability remains challenging. Low-level profiling tools (EAR~\cite{ear}, MERIC~\cite{meric}, LIKWID~\cite{likwid}) offer detailed measurement capabilities but add workflow complexity. Consequently, modeling teams often rely on job-scheduler logs, which vary in sampling frequency and boundary scope across centers.

\subsubsection{Standardization and Dynamic Grid Signals}
Establishing a unified framework requires consistent definitions for subsystem boundaries, sampling intervals, and idle power allocation. In addition, current practices rely on static annual grid averages ($135\ \mathrm{g\ CO_2/kWh}$), which ignore short-term fluctuations in renewable energy. However, adding carbon-aware scheduling to HPC introduces a key challenge: because supercomputing centers operate under high capital investments and must maintain near $100\%$ utilization, delaying a job often leads to zero-sum displacement, where another queued job simply fills the slot during high-emission periods. Genuine net emission reductions therefore require shifting workloads geographically across international centers or applying dynamic power capping.

\subsection{Recommendations for CMIP7 and Beyond}
Based on our empirical findings and literature analysis, we propose three recommendations for standardizing energy reporting in CMIP7:
\begin{enumerate}
    \item \textbf{Mandate Multi-Boundary Reporting:} CMIP7 should establish \textit{Operational Carbon} as the core baseline reporting metric while recommending \textit{Embodied Carbon} to be reported as an additive Scope 3 component.
    \item \textbf{Automate Scheduler-Level Measurement:} HPC facility operators should integrate automated energy measurement into batch schedulers (e.g., Slurm) to capture metered node power and idle baselines without requiring manual code instrumentation.
    \item \textbf{Incentivize Geographic and Carbon-Aware Scheduling:} Supercomputing centers and climate consortia should leverage dynamic grid carbon intensity signals to coordinate spatial shifting of flexible ensemble workloads across distributed international facilities and explore carbon-aware power capping.
\end{enumerate}

\section{Conclusions}
As climate modeling enters the kilometer-scale regime, evaluating performance solely by throughput metrics such as Simulated Years Per Day ($\mathrm{SYPD}$) or total core-hours is no longer sufficient. Environmental footprint accounting must be integrated into standard benchmarking practices. Building upon the voluntary reporting in CMIP6, the CMIP7 Task Team on Energy Consumption provides a framework for standardized pre-execution energy profiling across global HPC facilities.

Our evaluation of Destination Earth on MareNostrum 5 demonstrates that reported footprints depend heavily on boundary selection. Active-only accounting ($5.81\text{ MWh/SY}$, $847.7\ \mathrm{kg\kern 0.1em CO_2e/SY}$) isolates dynamic kernel efficiency, whereas operational accounting ($10.3\text{ MWh/SY}$, $1,501.7\ \mathrm{kg\kern 0.1em CO_2e/SY}$) reflects total facility utility demand. Furthermore, full life-cycle accounting—adding manufacturing emissions ($967.4\ \mathrm{kg\kern 0.1em CO_2e/SY}$) and TCO (€8,664/SY)—increases the total footprint by $1.64\times$ to $2,469.1\ \mathrm{kg\kern 0.1em CO_2e/SY}$, showing that Scope~3 embodied debt and capital costs are major factors.

Addressing current energy measurement challenges requires balancing low-level hardware profiling precision against job-scheduler usability, as well as transitioning from static annual grid averages to dynamic real-time carbon tracking.

To advance energy accountability, we advocate a three-part strategy for CMIP7: (1) mandating operational carbon as the baseline metric with embodied carbon as an additive Scope 3 multiplier, (2) automating energy measurement within batch schedulers, and (3) adopting carbon-aware scheduling for ensemble simulations. Standardizing these methodologies will help embed energy and carbon efficiency into future HPC system design and climate modeling practices.

\begin{acks}
This work has received funding from the European High Performance Computing Joint Undertaking (EuroHPC JU) and the European Union through the ESiWACE3 project under grant agreement No.~101093054. This work was also supported by the Destination Earth (DestinE) initiative of the European Commission. The authors also gratefully acknowledge the computational resources and support provided by the Barcelona Supercomputing Center (BSC) on the MareNostrum 5 supercomputer.
\end{acks}

\nobalance
\bibliographystyle{ACM-Reference-Format}
\Urlmuskip=0mu plus 1mu\relax
{\sloppy\raggedright\bibliography{mainbib}}

@inproceedings{powerwall,
  author    = {Villa, Oreste and Johnson, Daniel R. and Oconnor, Mike and Bolotin, Evgeny and Nellans, David and Luitjens, Justin and Sakharnykh, Nikolai and Wang, Peng and Micikevicius, Paulius and Scudiero, Anthony and Keckler, Stephen W. and Dally, William J.},
  booktitle = {SC '14: Proceedings of the International Conference for High Performance Computing, Networking, Storage and Analysis},
  title     = {Scaling the Power Wall: A Path to Exascale},
  year      = {2014},
  volume    = {},
  number    = {},
  pages     = {830-841},
  doi       = {10.1109/SC.2014.73}
}

@article{cmip6-cpmips,
  author  = {Acosta, M. C. and Palomas, S. and Paronuzzi Ticco, S. V. and Utrera, G. and Biercamp, J. and Bretonniere, P.-A. and Budich, R. and Castrillo, M. and Caubel, A. and Doblas-Reyes, F. and Epicoco, I. and Fladrich, U. and Joussaume, S. and Kumar Gupta, A. and Lawrence, B. and Le Sager, P. and Lister, G. and Moine, M.-P. and Rioual, J.-C. and Valcke, S. and Zadeh, N. and Balaji, V.},
  title   = {The computational and energy cost of simulation and storage for climate science:
             lessons from CMIP6},
  journal = {Geoscientific Model Development},
  volume  = {17},
  year    = {2024},
  number  = {8},
  pages   = {3081--3098},
  url     = {https://gmd.copernicus.org/articles/17/3081/2024/},
  doi     = {10.5194/gmd-17-3081-2024}
}

@misc{cmip7-taskteam-energy,
  author = {{World Climate Research Programme (WCRP)}},
  title  = {{CMIP7 Task Team: Energy Consumption}},
  year   = {2026},
  url    = {https://wcrp-cmip.org/cmip7-task-teams/energy-consumption/},
  note   = {Accessed: 2026-06-30}
}

@article{climatedt,
  author  = {Doblas-Reyes, F. J. and Kontkanen, J. and Sandu, I. and Acosta, M. and Al Turjmam, M. H. and Alsina-Ferrer, I. and Andr\'es-Mart\'{\i}nez, M. and Anerdi, C. and Arriola, L. and Axness, M. and Batlle Mart\'{\i}n, M. and Bauer, P. and Becker, T. and Beltr\'an, D. and Beyer, S. and Bockelmann, H. and Bretonni\`ere, P.-A. and Cabaniols, S. and Caprioli, S. and Castrillo, M. and Chandrasekar, A. and Cheedela, S. and Correal, V. and Danovaro, E. and Davini, P. and Enkovaara, J. and Frauen, C. and Fr\"uh, B. and Gaya \`Avila, A. and Ghinassi, P. and Ghosh, R. and Ghosh, S. and Gonz\'alez, I. and Grayson, K. and Griffith, M. and Hadade, I. and Haine, C. and Hartick, C. and Haus, U.-U. and Hearne, S. and J\"arvinen, H. and Jim\'enez, B. and John, A. and Juchem, M. and Jung, T. and Kegel, J. and Kelbling, M. and Keller, K. and Kinoshita, B. and Kiszler, T. and Klocke, D. and Kluft, L. and Koldunov, N. and K\"olling, T. and Kolstela, J. and Kornblueh, L. and Kosukhin, S. and Lacima-Nadolnik, A. and Leal Rojas, J. J. and Lehtiranta, J. and Lunttila, T. and Luoma, A. and Manninen, P. and Medvedev, A. and Milinski, S. and Mohammed, A. and M\"uller, S. and Naryanappa, D. and Nazarova, N. and Niemel\"a, S. and Niraula, B. and Nortamo, H. and Nummelin, A. and Nurisso, M. and Ortega, P. and Paronuzzi, S. and Pedruzo-Bagazgoitia, X. and Pelletier, C. and Pe\~na, C. and Polade, S. and Pradhan, H. K. and Quintanilla, R. and Quintino, T. and Rackow, T. and R\"ais\"anen, J. and Rajput, M. M. and Redler, R. and Reuter, B. and Rocha Monteiro, N. and Roura-Adserias, F. and Ruppert, S. and Sayed, S. and Schnur, R. and Sharma, T. and Sidorenko, D. and Sievi-Korte, O. and Soret, A. and Steger, C. and Stevens, B. and Streffing, J. and Sunny, J. and Tenorio, L. and Thober, S. and Tigerstedt, U. and Tinto, O. and Tonttila, J. and Tuomenvirta, H. and Tuppi, L. and Van Thielen, G. and Vitali, E. and von Hardenberg, J. and Wagner, I. and Wedi, N. and Wehner, J. and Willner, S. and Yepes-Arb\'os, X. and Ziemen, F. and Zimmermann, J.},
  title   = {The Destination Earth digital twin for climate change adaptation},
  journal = {Geoscientific Model Development},
  volume  = {19},
  year    = {2026},
  number  = {7},
  pages   = {2821--2848},
  url     = {https://gmd.copernicus.org/articles/19/2821/2026/},
  doi     = {10.5194/gmd-19-2821-2026}
}

@inproceedings{ear,
  author    = {Corbalan, Julita and Brochard, Luigi and Lakhlili, Jalal and d'Amico, Marco and Vidal, Oriol and Aneas, Jordi},
  title     = {Energy Data Center Monitoring and Management with EAR},
  year      = {2025},
  isbn      = {9798400711251},
  publisher = {Association for Computing Machinery},
  address   = {New York, NY, USA},
  url       = {https://doi.org/10.1145/3679240.3735105},
  doi       = {10.1145/3679240.3735105},
  booktitle = {Proceedings of the 16th ACM International Conference on Future and Sustainable Energy Systems},
  pages     = {915–920},
  numpages  = {6},
  location  = {
               },
  series    = {E-Energy '25}
}

@inproceedings{meric,
  author    = {Vysocky, Ondrej
               and Beseda, Martin
               and {\v{R}}{\'i}ha, Lubom{\'i}r
               and Zapletal, Jan
               and Lysaght, Michael
               and Kannan, Venkatesh},
  editor    = {Kozubek, Tom{\'a}{\v{s}}
               and {\v{C}}erm{\'a}k, Martin
               and Tich{\'y}, Petr
               and Blaheta, Radim
               and {\v{S}}{\'i}stek, Jakub
               and Luk{\'a}{\v{s}}, Dalibor
               and Jaro{\v{s}}, Ji{\v{r}}{\'i}},
  title     = {MERIC and RADAR Generator: Tools for Energy Evaluation and Runtime Tuning of HPC Applications},
  booktitle = {High Performance Computing in Science and Engineering},
  year      = {2018},
  publisher = {Springer International Publishing},
  address   = {Cham},
  pages     = {144--159},
  isbn      = {978-3-319-97136-0}
}

@inproceedings{likwid,
  title        = {Energy efficient frequency scaling on GPUs in heterogeneous HPC systems},
  author       = {Kraljic, Karlo and Kerger, Daniel and Schulz, Martin},
  booktitle    = {International Conference on Architecture of Computing Systems},
  pages        = {3--16},
  year         = {2022},
  organization = {Springer}
}

@inproceedings{hpc-co2,
  author    = {Rao, Varsha and Chien, Andrew A.},
  title     = {Modeling the Carbon Footprint of HPC: The Top 500 and EasyC},
  year      = {2025},
  isbn      = {9798400718717},
  publisher = {Association for Computing Machinery},
  address   = {New York, NY, USA},
  url       = {https://doi.org/10.1145/3731599.3767567},
  doi       = {10.1145/3731599.3767567},
  booktitle = {Proceedings of the SC '25 Workshops of the International Conference for High Performance Computing, Networking, Storage and Analysis},
  pages     = {2032–2040},
  numpages  = {9},
  location  = {
               },
  series    = {SC Workshops '25}
}

@article{eyring2016cmip6,
  author  = {Eyring, V. and Bony, S. and Meehl, G. A. and Senior, C. A. and Stevens, B. and Stouffer, R. J. and Taylor, K. E.},
  title   = {Overview of the Coupled Model Intercomparison Project Phase 6 (CMIP6)
             experimental design and organization},
  journal = {Geoscientific Model Development},
  volume  = {9},
  year    = {2016},
  number  = {5},
  pages   = {1937--1958},
  url     = {https://gmd.copernicus.org/articles/9/1937/2016/}
}

@inproceedings{strubell-etal-2019-energy,
  title     = {Energy and Policy Considerations for Deep Learning in {NLP}},
  author    = {Strubell, Emma  and
               Ganesh, Ananya  and
               McCallum, Andrew},
  editor    = {Korhonen, Anna  and
               Traum, David  and
               M{\`a}rquez, Llu{\'i}s},
  booktitle = {Proceedings of the 57th Annual Meeting of the Association for Computational Linguistics},
  month     = jul,
  year      = {2019},
  address   = {Florence, Italy},
  publisher = {Association for Computational Linguistics},
  url       = {https://aclanthology.org/P19-1355/},
  doi       = {10.18653/v1/P19-1355},
  pages     = {3645--3650}
}

@article{patterson2021carbon,
  author     = {David A. Patterson and
                Joseph Gonzalez and
                Quoc V. Le and
                Chen Liang and
                Lluis{-}Miquel Munguia and
                Daniel Rothchild and
                David R. So and
                Maud Texier and
                Jeff Dean},
  title      = {Carbon Emissions and Large Neural Network Training},
  journal    = {CoRR},
  volume     = {abs/2104.10350},
  year       = {2021},
  url        = {https://arxiv.org/abs/2104.10350},
  eprinttype = {arXiv},
  eprint     = {2104.10350},
  bibsource  = {dblp computer science bibliography, https://dblp.org}
}

@article{wu2022sustainable,
  author     = {Carole-Jean Wu and Ramya Raghavendra and Udit Gupta and Bilge Acun and Newsha Ardalani and Kiwan Maeng and Gloria Chang and Fiona Aga Behram and James Huang and Charles Bai and Michael Gschwind and Anurag Gupta and Myle Ott and Anastasia Melnikov and Salvatore Candido and David Brooks and Geeta Chauhan and Benjamin C. Lee and Hsien-Hsin S. Lee and Bugra Akyildiz and Maximilian Balandat and Joe Spisak and Ravi Jain and Mike Rabbat and Kim M. Hazelwood},
  title      = {Sustainable {AI:} Environmental Implications, Challenges and Opportunities},
  journal    = {CoRR},
  volume     = {abs/2111.00364},
  year       = {2021},
  url        = {https://arxiv.org/abs/2111.00364},
  eprinttype = {arXiv},
  eprint     = {2111.00364},
  bibsource  = {dblp computer science bibliography, https://dblp.org}
}

@inproceedings{10.1145/3581784.3607035,
  author    = {Li, Baolin and Basu Roy, Rohan and Wang, Daniel and Samsi, Siddharth and Gadepally, Vijay and Tiwari, Devesh},
  title     = {Toward Sustainable HPC: Carbon Footprint Estimation and Environmental Implications of HPC Systems},
  year      = {2023},
  isbn      = {9798400701092},
  publisher = {Association for Computing Machinery},
  address   = {New York, NY, USA},
  url       = {https://doi.org/10.1145/3581784.3607035},
  doi       = {10.1145/3581784.3607035},
  booktitle = {Proceedings of the International Conference for High Performance Computing, Networking, Storage and Analysis},
  articleno = {19},
  numpages  = {15},
  location  = {Denver, CO, USA},
  series    = {SC '23}
}

@inproceedings{10495563,
  author    = {Masciari, Elio and Napolitano, Enea Vincenzo},
  booktitle = {2024 32nd Euromicro International Conference on Parallel, Distributed and Network-Based Processing (PDP)},
  title     = {The Environmental Cost of High Performance Computing System Simulation},
  year      = {2024},
  volume    = {},
  number    = {},
  pages     = {289-292},
  doi       = {10.1109/PDP62718.2024.00048}
}

@manual{acosta_2025_17522934,
  title  = {Metrics to quantify performance, energy and {CO$_2$} footprint of {CMIP7} simulations},
  author = {Acosta, Mario C. and
            Palomas, Sergi and
            Valcke, Sophie and
            Bretonni{\`e}re, Pierre-Antoine and
            Amjad, Muhammad and
            Mahmood, Asad and
            Smith, Paul and
            Hassell, David and
            Bonou, Fr{\'e}d{\'e}ric},
  month  = nov,
  year   = 2025,
  doi    = {10.5281/zenodo.17522934},
  url    = {https://doi.org/10.5281/zenodo.17522934}
}

@misc{electricitymaps,
  author = {{Electricity Maps}},
  title  = {Spain Electricity Grid Carbon Intensity Data and Methodology},
  year   = {2025},
  url    = {https://app.electricitymaps.com/map/zone/ES/all/yearly},
  note   = {Accessed: 2026}
}

@misc{eurohpc-mn5,
  author = {{European High Performance Computing Joint Undertaking (EuroHPC JU)}},
  title  = {MareNostrum 5 Supercomputer Inauguration and System Specifications},
  year   = {2023},
  url    = {https://eurohpc-ju.europa.eu/supercomputers/marenostrum-5_en},
  note   = {Accessed: 2026}
}

\end{document}